\documentclass[useAMS,usenatbib,usegraphicx]{mn2e}

\title[TTV and activity in the WASP-10 planetary system]{Transit timing variation and activity in the WASP-10 planetary system\thanks{Partly based on observations made with the 0.6 and 2.0-m telescopes of the Rozhen National Astronomical Observatory, which is operated by the Institute of Astronomy, Bulgarian Academy of Sciences and the 90-cm telescope of the University Observatory Jena, which is operated by the Astrophysical Institute of the Friedrich Schiller University.}}
\author[G. Maciejewski et al.]
{G.~Maciejewski,$^{1,2}$\thanks{E-mail: gm@astro.uni-jena.de}
D.~Dimitrov,$^{3}$ 
R.~Neuh\"auser,$^{1}$ 
N.~Tetzlaff,$^{1}$ 
A.~Niedzielski,$^{2}$
\newauthor
St.~Raetz,$^{1}$ 
W.~P.~Chen,$^{4}$
F.~Walter,$^{5}$
C.~Marka,$^{1}$ 
S.~Baar,$^{1}$
T.~Krejcov\'a,$^{6}$
\newauthor
J.~Budaj,$^{7}$
V.~Krushevska,$^{7,8}$
K.~Tachihara,$^{9}$
H.~Takahashi$^{10}$
and M.~Mugrauer$^{1}$
\\
$^{1}$Astrophysikalisches Institut und Universit\"ats-Sternwarte, Schillerg\"asschen 2--3, D--07745 Jena, Germany\\
$^{2}$Toru\'n Centre for Astronomy, N. Copernicus University, Gagarina 11, PL--87100 Toru\'n, Poland\\
$^{3}$Institute of Astronomy, Bulgarian Academy of Sciences, 72 Tsarigradsko Chausse Blvd., 1784 Sofia, Bulgaria\\
$^{4}$Institute of Astronomy, National Central University, 300 Jhongda Rd., Jhongli 32001, Taiwan\\
$^{5}$Department of Physics and Astronomy, SUNY, Stony Brook, NY 11794-3800, USA\\
$^{6}$Masaryk University, Department of Theoretical Physics and Astrophysics, 602 00 Brno, Czech Republic\\
$^{7}$Astronomical Institute of the Slovak Academy of Sciencies, 059 60 Tatransk\'a Lomnica, Slovak Republic\\
$^{8}$Main Astronomical Observatory of National Academy of Sciences of Ukraine, 27 Akademika Zabolotnoho St., 03680 Kyiv, Ukraine\\
$^{9}$National Astronomical Observatory of Japan, ALMA project office, 2-21-1 Osawa Mitaka Tokyo 181-8588 Japan\\
$^{10}$Gunma Astronomical Observatory, 6860-86 Nakayama, Takayama-mura, Agatsuma-gun, Gunma 377-0702 Japan}
\begin{document}

\date{Accepted. Received ; in original form }

\pagerange{\pageref{firstpage}--\pageref{lastpage}} \pubyear{2010}

\maketitle

\label{firstpage}

\begin{abstract}
Transit timing analysis may be an effective method of discovering additional bodies in extrasolar systems which harbour transiting exoplanets. The deviations from the Keplerian motion, caused by mutual gravitational interactions between planets, are expected to generate transit timing variations of transiting exoplanets. In 2009 we collected 9 light curves of 8 transits of the exoplanet WASP-10b. Combining these data with published ones, we found that transit timing cannot be explained by a constant period but by a periodic variation. Simplified three-body models which reproduce the observed variations of timing residuals were identified by numerical simulations. We found that the configuration with an additional planet of mass of $\sim$0.1 $M_{\rm{J}}$ and orbital period of $\sim$5.23 d, located close to the outer 5:3 mean motion resonance, is the most likely scenario. If the second planet is a transiter, the estimated flux drop will be $\sim$0.3 per cent and can be observable with a ground-based telescope. Moreover, we present evidence that the spots on the stellar surface and rotation of the star affect the radial velocity curve giving rise to spurious eccentricity of the orbit of the first planet. We argue that the orbit of WASP-10b is essentially circular. Using the gyrochronology method, the host star was found to be $270 \pm 80$ Myr old. This young age can explain the large radius reported for WASP-10b.
\end{abstract}

\begin{keywords}
planetary systems -- stars: individual: WASP-10.
\end{keywords}

\begin{table*}
\centering
\begin{minipage}{170mm}
\caption{List of telescopes: FoV is the field of view of the instrument and $N_{\rm{tr}}$ is the number of observed transits.} 
\label{tabela1}
\begin{tabular}{c l l l c}
\hline
ID & Telescope & Observatory & Detector &  $N_{\rm{tr}}$ \\
   &           & Location    & CCD size & Field of view   \\
\hline 
1 & 0.6-m Cassegrain & Astronomical Observatory, N.Copernicus University & SBIG STL-1001 & 1 \\
  &                  & Piwnice near Toru\'n, Poland & $1024 \times 1024$, 24 $\mu$m & $11\farcm8 \times 11\farcm8$  \\
2 & 0.6-m Schmidt$^*$    & University Observatory Jena,  & CCD-imager STK & 2 \\
  &                  & Gro{\ss}schwabhausen near Jena, Germany & $2048\times 2048$, 13.5 $\mu$m & $52\farcm8 \times 52\farcm8$  \\
3 & 0.6-m Cassegrain & National Astronomical Observatory & FLI PL09000 & 4 \\
  &                  & Rozhen, Bulgaria & $3056 \times 3056$, 12$\mu$m & $17\farcm3 \times 17\farcm3$  \\
4 & 2-m Ritchey-Chr\'etien & National Astronomical Observatory & PI VersArray:1300B & 1 \\
  &                  & Rozhen, Bulgaria & $1340 \times 1300$, 20$\mu$m & $5\farcm8 \times 5\farcm6$  \\
5 & 0.8-m Tenagra II & Tenagra Observatories & SITe & 1 \\
  &                  & Arizona, USA & $1024 \times 1024$, 24$\mu$m & $14\farcm8 \times 14\farcm8$  \\
6 & 1.5-m Ritchey-Chr\'etien & Gunma Astronomical Observatory & Andor DW432 & 0$^{**}$ \\
  &                  & Takayama, Japan & $1250 \times 1152$, 24$\mu$m & $12\farcm5 \times 11\farcm5$  \\
\hline
\end{tabular}
\\$^*$~see \citet{mugrauer} for details
\\$^{**}$~no~scientific output due to bad weather conditions
\end{minipage}
\end{table*}

\section{Introduction}

Analysing transit timing variations (TTV) of exoplanets is expected to be an efficient method to discover additional low-mass planets \citep{miralda02, Schneider04, holmanmurray05, agoletal05, steffenetal07}. Many studies have been performed to detect TTV signals but they resulted only in constraints on the parameters of hypothetical second planets (e.g. \citealt{gibsonetal09a}). Recently, \citet{Maciejewski10} detected variation in the transit timing of WASP-3b. They found that a configuration with a hypothetical second planet of mass of $\sim$15 $M_{\earth}$, located close to an outer 2:1 mean motion resonance (MMR) may reproduce the observed TTV signal. \citet{Lendl} found some indications for the TTV for OGLE2-TR-L9b but no preliminary solution was proposed.

WASP-10b is an exoplanet discovered by \citet{Christian09}. The mass and radius of the planet were found to be $2.96^{+0.22}_{-0.17}$ $M_{\rmn{J}}$ and $1.28^{+0.08}_{-0.09}$ $R_{\rmn{J}}$, respectively, larger than interior models of irradiated giant planets predict. The planet orbits its host star in $\sim$3.09 d at an orbital semi-major axis of $0.0369^{+0.0012}_{-0.0014}$ au. The eccentricity of the orbit was reported to be $0.059^{+0.014}_{-0.004}$ -- a large value for a close-in planet. In the discovery paper \citet{Christian09} determined transit depth and duration to be 29 mmag and 2.36 h, respectively. 

\citet{johnson09} and \citet{johnson10} redetermined the planet's parameters with a high-precision light curve using the University of Hawaii 2.2-m telescope with a photometric precision of $4.7 \times 10^{-4}$ and the mid-transit time error of 7 s. The mass of the planet was found to be $3.15^{+0.13}_{-0.11}$ $M_{\rmn{J}}$ and the planetary radius turned out to be noticeably smaller, i.e. $1.08 \pm 0.02$ $R_{\rmn{J}}$. The latter result was not confirmed by the studies of \citet*{Krejcova10} or \citet{Dittmann10}. The first team found the radius of WASP-10b to be in excellent agreement with the value obtained by \citet{Christian09}, and a similar conclusion was reached by \citet{Dittmann10} with no TTV signature detected. 

The host star ($V=12.7$ mag) is a K5 dwarf with effective temperature of $4675 \pm 100$ K, located $90 \pm 20$ pc from the Sun \citep{Christian09}. \citet{Smith09} confirmed previous evidences that the light curve of WASP-10 exhibits a rotational photometric variability caused by star spots. The stellar rotation period was found to be $11.91\pm0.05$ d -- in agreement with $\sim$12 d reported by \citet{Christian09}. 

The presence of an additional small planet was suggested to pump the eccentricity of WASP-10b \citep{Christian09}. This system was therefore a good target for TTV studies.

The order of this paper is as follows. In Section 2 observations and data reduction are summarised. The discovery of the periodic transit timing variation is presented in Section 3. The process of identifying and verifying of the most likely 2-planetary model of the WASP-10 system is described in Section 4. A discussion of the proposed scenario is presented in Section 5. Finally, conclusions are collected in Section 6.

\section[]{Observations and data reduction}

We collected 9 light curves of 8 transits of WASP-10b during a dedicated international observing campaign involving telescopes spread worldwide at different longitudes (listed in Table~\ref{tabela1} and \ref{tabela2}). The transit on July 31, 2009 was observed with two instruments in different observatories. The telescope diameters of 0.6 to 2.0 m allowed us to collect photometric data with $1.1 - 2.7$ mmag precision. Observations generally started $\sim$1 hour before the expected beginning of a transit and ended $\sim$1 hour after the event. However, weather conditions and schedule constraints meant that we were not always able to follow this scheme.

Standard IDL procedures (adapted from DAOPHOT) were used for the reduction of the photometric data collected at the National Astronomical Observatory at Rozhen, Bulgaria, and computing the differential aperture photometry. Using the method of \citet{everetthowell01} several stars (4 to 6) with photometric precision better than 5 mmag were selected to create an artificial standard star used for differential photometry. Data from the remaining telescopes were reduced with the software pipeline developed for the Semi-Automatic Variability Search sky survey \citep*{niedzielskietal03}. To generate an artificial comparison star, 30--50 per cent of the stars with the lowest light-curve scatter were selected iteratively from the field stars brighter than 3 mag below the saturation level. To measure instrumental magnitudes, various aperture radii were used. The aperture which was found to produce light curves with the smallest scatter was used to generate a final light curve.

\begin{table}
\centering
\caption{The summary of observing runs: ID -- the identification of the instrument according to Table~\ref{tabela1}, $N_{\rmn{exp}}$ -- number of useful exposures, $T_{\rmn{exp}}$ -- exposure times. Dates UT are given for the beginning of nights.} 
\label{tabela2}
\begin{tabular}{c l c c c c c}
\hline
Run & Date & ID & Filter & $N_{\rmn{exp}}$ & $T_{\rmn{exp}}$ (s)\\
\hline 
1 & 2009 Jan. 5  & 1 & $R$ & $254$ & 50 \\
2 & 2009 July 31 & 2 & $R$ & $144$ & 60, 80  \\
3 & 2009 July 31 & 3 & $R$ & $116$ & 120  \\
4 & 2009 Aug. 3  & 3 & $R$ & $122$ & 90  \\
5 & 2009 Aug. 28 & 3 & $R$ &  $99$ & 90  \\
6 & 2009 Aug. 31 & 2 & $R$ & $203$ & 50, 60  \\
7 & 2009 Sep. 3  & 3 & $R$ & $105$ & 90  \\
8 & 2009 Oct. 1  & 4 & $V$ & $548$ & 15  \\
9 & 2009 Nov. 17 & 5 & $R$ &  $76$ & 90 \\
\hline
\end{tabular}
\end{table}

\begin{figure*}
  \includegraphics[width=17cm]{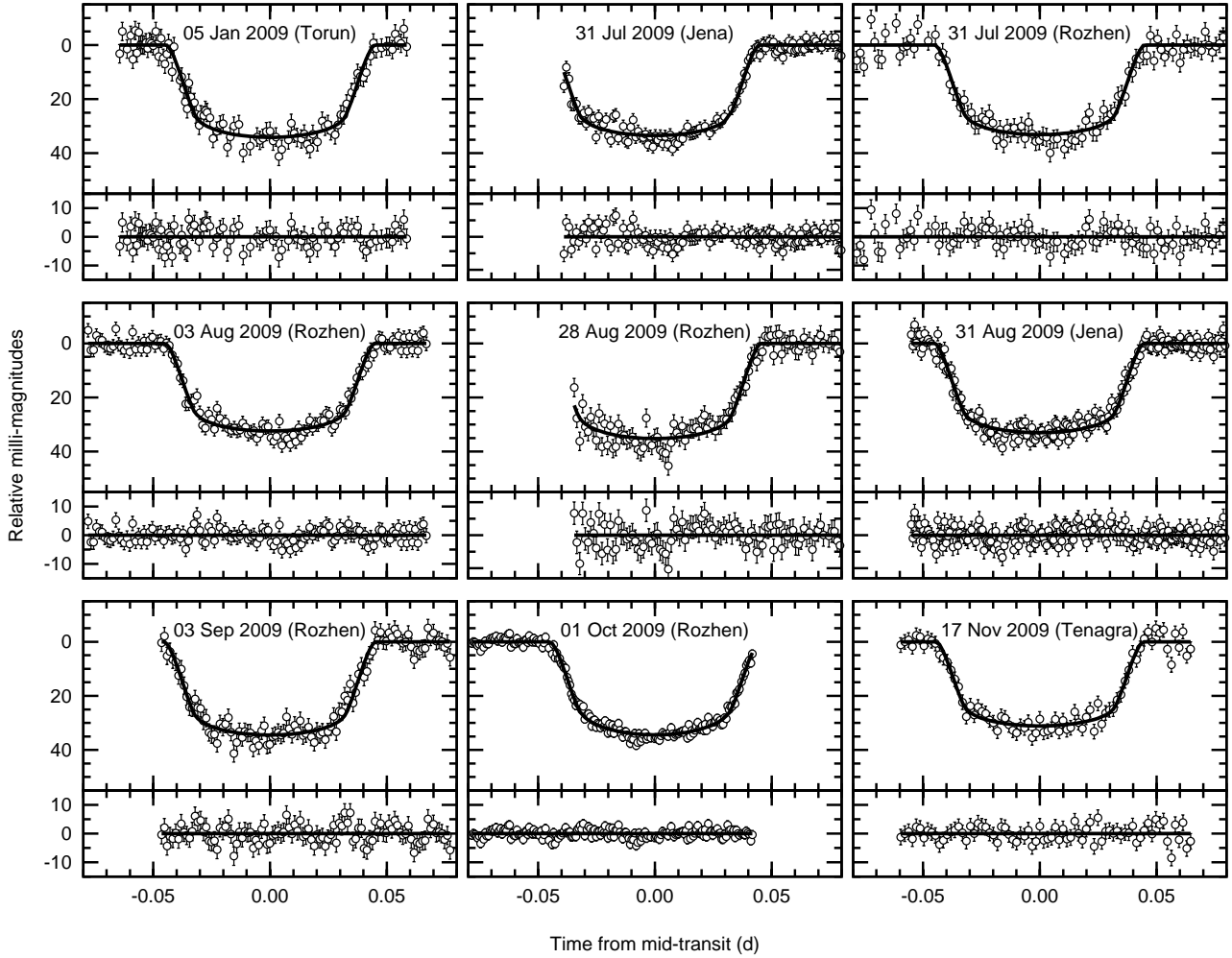}
  \caption{Light curves of WASP-10b transits in individual runs. The best-fitting models are shown as continuous lines. The transit on 31 July was observed simultaneously by two observatories.}
  \label{rys1}
\end{figure*}

\begin{figure*}
 \includegraphics[width=17cm]{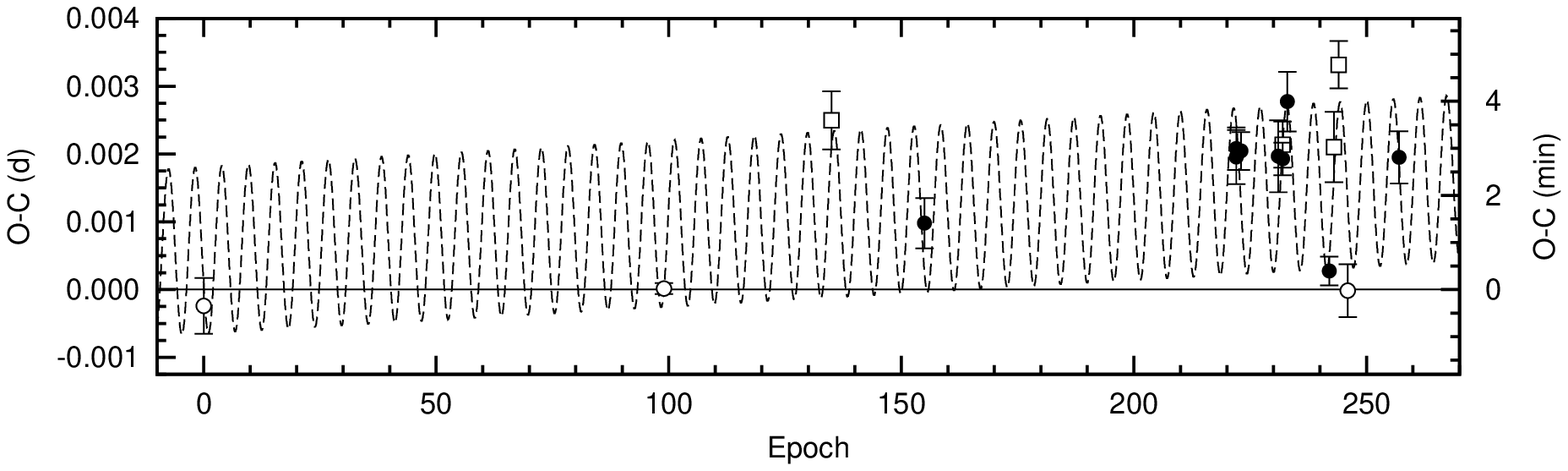}
 \caption{The observation minus calculation ($O-C$) diagram for WASP-10b generated for linear ephemeris based on literature data (continuous line). Open circles denote literature data taken from \citet{Christian09}, \citet{johnson10} and \citet{Dittmann10}. Open squares are based on reanalysed photometry from \citet{Krejcova10}. Filled symbols denote results from our campaign. A significantly better fit may be obtained for the new ephemeris given by Eq.~1 (dashed line). $3\sigma$ error bars were taken for mid-transit time reported by \citet{Dittmann10}. Their light curve was not corrected for a linear trend which is well visible in out-of-transit phases (J. Dittmann 2010, priv. comm.). Our tests showed that this effect significantly affected the accuracy of the mid-transit time.}
 \label{rys2}
\end{figure*}

\section[]{Results}
 
\subsection[]{Light curve analysis} 

A model-fitting algorithm available via the Exoplanet Transit Database \citep*{poddanyetal10} was used to derive transit parameters: duration, depth and mid-transit time. The procedure employs the \textsc{occultsmall} routine of \citet{mandelalgol02} and the Levenberg--Marquardt non-linear least squares fitting algorithm which also provides uncertainties. As our research was focused on determining mid-transit times, the number of parameters to be fitted could be reduced. An impact parameter $b=a \cos i / R_{*} = 0.299^{+0.029}_{-0.043}$, where $a$ is semi-major axis, $i$ is inclination and $R_{*}$ is host-star radius, was taken from \citet{johnson09} and was fixed during the fitting procedure. We used the linear limb-darkening law by \citet{vanhamme93} with the limb-darkening coefficient linearly interpolated for the host star in a given filter. A first- or second-order polynomial was used to remove trends in magnitude or flux. The mid-transit times were corrected from UTC to Terrestrial Dynamical Time (TT) and then transformed into BJD \citep*{Eastman10}. Light curves with best-fitting models and residuals are shown in Fig.~\ref{rys1} and derived parameters are collected in Table~\ref{tabela3}.

\citet{Krejcova10} also published observations and $R$-band light curves of four complete transit events of WASP-10b. In their analysis they focussed mainly on the determination of the planet radius and did not study the mid-transit times, $O-C$ diagram or TTV. We determine 4 mid-transit times for these observations and placed them into the context of our data to create a larger sample for TTV analysis. We used the same method to determine the mid-transit times from their light curves as for all other observations presented here. Corresponding, mid-transit times are presented in Table~\ref{tabela3}.

Across all data sets, two transits on epochs 222 and 232 were observed simultaneously by two different telescopes. The data were reduced with different pipelines and by different teams. The transit on epoch 222 was observed with 60-cm telescopes in Jena and at Rozhen. The difference in mid-transit times was found to be $\sim$11 s. The transit on epoch 232 was covered by observations in Jena and the 50-cm telescope in Star\'a Lesn\'a in Slovakia \citep{Krejcova10}. In this case the discrepancy of timing was found to be $\sim$18 s. Timing differences are well within error bars of individual determinations for both transits. This practical test strengthens the reliability of our mid-transit time determinations.

\begin{table*}
\centering
\begin{minipage}{155mm}
\caption{Parameters of transit light-curve modelling. $T_{0}$  denotes the mid-transit time, $T_{\rmn{d}}$ -- transit time duration, $\delta$ -- depth, $\sigma$ -- averaged standard deviation of the fit and $E$ -- epoch. $O-C$ values were calculated according to the linear part of Eq.~\ref{rownanie1} and the best-fitting parameters. $O-C$ values are given both in days and in units of multiplies of the uncertainty of the mid-transit times. $t_{\rmn{cad}}$ is the mean cadence of a light curve. Results of reanalysed data from \citet{Krejcova10} are collected below the horizontal line. BJD times are based on Terrestrial Dynamical Time (TT).} 
\label{tabela3}
\begin{tabular}{c c c c c c c c c}
\hline
Run & $T_{0}$ & $T_{\rmn{d}}$ & $\delta$ & $\sigma$ & $E$ & $O-C$ &  $O-C$ & $t_{\rmn{cad}}$ \\
    & $\rmn{BJD_{TT}}-2450000$  & (min)   & (mmag)   & (mmag)   &     & (d) & ($T_{0}$ errors) & (s)\\
\hline 
1 & $4837.23116\pm0.00037$ & $128.9\pm1.4$ & $34.1\pm0.8$ & 2.7 & 155 & $-0.00030$ & $-0.8$ & 56\\
2 & $5044.44427\pm0.00032$ & $128.9\pm1.2$ & $33.5\pm1.1$ & 1.7 & 222 & $+0.00069$ & $+2.2$ & 87\\
3 & $5044.44415\pm0.00040$ & $129.1\pm1.5$ & $33.1\pm0.7$ & 2.6 & 222 & $+0.00057$ & $+1.4$ & 130\\
4 & $5047.53696\pm0.00028$ & $129.4\pm1.0$ & $32.4\pm0.5$ & 2.0 & 223 & $+0.00066$ & $+2.3$ & 103\\
5 & $5072.27861\pm0.00053$ & $131.1\pm2.1$ & $35.2\pm0.8$ & 2.5 & 231 & $+0.00056$ & $+1.1$ & 102\\
6 & $5075.37129\pm0.00024$ & $129.2\pm0.8$ & $33.0\pm0.7$ & 2.0 & 232 & $+0.00053$ & $+2.2$ & 63\\
7 & $5078.46485\pm0.00044$ & $130.4\pm1.5$ & $34.5\pm1.1$ & 2.5 & 233 & $+0.00137$ & $+3.1$ & 102 \\
8 & $5106.29681\pm0.00021$ & $129.8\pm0.7$ & $34.4\pm0.7$ & 1.1 & 242 & $-0.00115$ & $-5.4$ & 19\\
9 & $5152.68923\pm0.00039$ & $128.8\pm1.3$ & $33.3\pm2.4$ & 2.1 & 257 & $+0.00051$ & $+1.3$ & 143\\
\hline
  & $4775.37834\pm0.00042$&   &   &  & 135 & $+0.00125$ & $+2.9$ & 35\\
  & $5075.37150\pm0.00034$&   &   &  & 232 & $+0.00073$ & $+2.2$ & 26\\
  & $5109.39135\pm0.00052$&   &   &  & 243 & $+0.00068$ & $+1.3$ & 36\\
  & $5112.48528\pm0.00035$&   &   &  & 244 & $+0.00189$ & $+5.4$ & 36\\
\hline
\end{tabular}
\end{minipage}
\end{table*}

\subsection[]{Detection of a TTV} 

\begin{figure}
 \includegraphics[width=84mm]{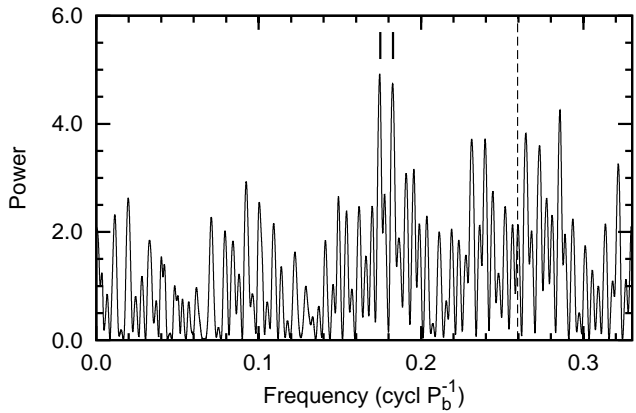}
 \caption{The Lomb--Scargle periodogram generated for timing residuals plotted in Fig.~\ref{rys2}, showing the existence of a periodic signal. The most significant peaks are indicated. The upper limit of the periodogram is determined by the Nyquist frequency for the data set. The vertical dashed line marks the rotational period of the host star (see Section 5).}
 \label{rys3}
\end{figure}

While determining new linear ephemeris, we noted that a linear fit of the epoch $E$ and orbital period $P_{\rm{b}}$ of WASP-10b resulted in reduced $\chi^2_{\rm{red}}=13.8$. Individual mid-transit errors were taken as weights in the fitting procedure. Such a high value of $\chi^2_{\rm{red}}$ suggests the existence of additional signal in the $O-C$ diagram (Fig.~\ref{rys2}). The Lomb--Scargle periodogram \citep{Lomb,Scargle} of the residuals revealed the existence of 2 peaks of similar significance and frequencies of 0.175 and 0.183 cycl~$P_{\rm{b}}^{-1}$ (Fig.~\ref{rys3}). For each frequency, an ephemeris was refitted with the linear trend plus a sinusoidal variation in the form:
\begin{equation}
	T_{\rmn{b}} = T_0 + E \times P_{\rm{b}} + A_{\rm{ttv}}\sin \left( 2 \pi \frac{E-E_{\rm{ttv}}}{P^{\rm{ttv}}} \right)
	\label{rownanie1}
\end{equation}
where $T_0$ is the mid-transit time for initial epoch ($E=0$), $A_{\rm{ttv}}$ is the semi-amplitude of detected transit timing variation, $E_{\rm{ttv}}$ is the epoch offset of the TTV signal and $P^{\rm{ttv}}$ is its period. The minimal $\chi^2_{\rm{red}}=3.2$ was obtained for $f^{\rmn{ttv}}_1=0.183$ cycl~$P_{\rm{b}}^{-1}$ and resulted in $T_0=2454357.86011 \pm 0.00047$ $\rm{BJD_{TT}}$, $P_{\rm{b}} = 3.0927183 \pm 0.0000021$ d, $A_{\rm{ttv}} = 0.00120 \pm 0.00036$ d, $E_{\rm{ttv}} = 1.8 \pm 0.4$ and $P^{\rm{ttv}}_1 = 5.473 \pm 0.012$ $P_{\rm{b}}$.  
    
The procedure run for $f^{\rmn{ttv}}_2=0.175$ cycl~$P_{\rm{b}}^{-1}$ gave $\chi^2_{\rm{red}}=4.1$, $P^{\rm{ttv}}_2 = 5.7172 \pm 0.0082$ $P_{\rm{b}}$ and similar values of $T_0$ and $P_{\rm{b}}$. In this case the fit turned out to be poorer, thus we used the ephemeris based on parameters obtained for $P^{\rm{ttv}}_1$. 

The ephemeris calculated according to Eq.~1 (and $P^{\rm{ttv}}_1$) is plotted in Fig.~\ref{rys2} with a dashed line. The $O-C$ values which are used in further analysis were calculated according to the linear part of Eq.~1.

\begin{figure}
 \includegraphics[width=84mm]{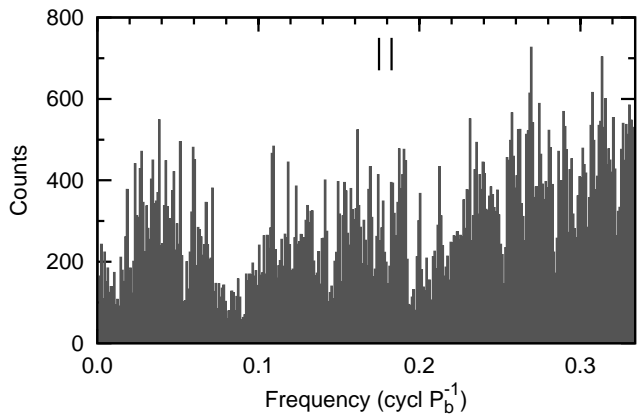}
 \caption{The distribution of dominant frequencies for $10^5$ fake $O-C$ diagrams with observed data points replaced by the white noise. The number of bins is equal to the square root of the number of data points in the sample and the width of individual bins is 0.001 cycl~$P_{\rm{b}}^{-1}$. Detected TTV frequencies, $f^{\rmn{ttv}}_1$ and $f^{\rmn{ttv}}_2$, are indicated. No significant peak is associated with any of them what indicates that a spectral window is not a source of detected periodicities.}
 \label{rys3b}
\end{figure}

To check if a random distribution of data points in the $O-C$ diagram may favour the detected frequencies, $10^5$ fake data sets were generated. The original residuals were replaced with random values with a white-noise distribution of the amplitude. Then, the Lomb--Scargle algorithm was run to find a dominant frequency in each fake diagram. The histogram of these frequencies, with the number of bins equal to  the square root of the number of data points in the sample, is presented in Fig.~\ref{rys3b}. Detected TTV frequencies, $f^{\rmn{ttv}}_1$ and $f^{\rmn{ttv}}_2$, are not associated with any significant peak in the histogram.  
 
\begin{figure}
 \includegraphics[width=84mm]{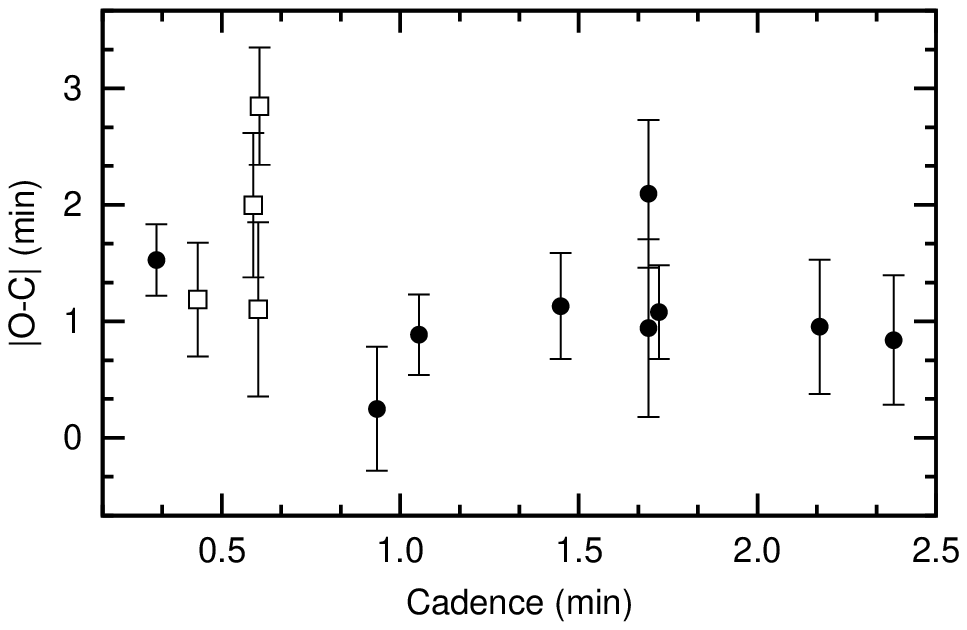}
 \caption{Absolute values of individual residuals in transit timing vs. cadence. Symbols denote the same data sets as in Fig.~\ref{rys2}. No correlation between both quantities was found.}
 \label{rys3c}
\end{figure}

\citet{KippingBakos10a} pointed out that the transit phasing which is the time difference between the expected mid-transit moment and the nearest data point in a light curve may generate a spurious TTV signal. For example, for a light curve of 60-s cadence time difference is expected to be $\pm 30$ s \citep{KippingBakos10b}. Cadences of our individual light curves are in a wide range between 19 and 143 s (see Table~\ref{tabela3}) with the median value of 63 s. This value is $\sim$3 times smaller than the amplitude of the detected TTV signal. As it is shown in Fig.~\ref{rys3c}, no correlation between individual residuals and cadences was detected. A formal least-square fit to this data set resulted in the correlation coefficient $R=-0.36$ which clearly reveals no correlation at the relevant level. The checks discussed above strengthen the detection of the TTV signal in our data set.  
      
\section[]{Two-planetary model}
 
The possible non-zero eccentricity postulated by \citet{Christian09} might potentially have resulted from confusion with a two-planet system in which both planets orbit their host star in circular orbits in an inner 2:1 resonance \citep{angladaetal10}. Although we argue later that eccentricity of WASP-10b is indistinguishable from zero, the case of an inner perturber was considered for completeness of our analysis\footnote{We do not analyse the scenario with an exomoon orbiting WASP-10b because no transit duration variation, predicted by \citet{kipping09}, was detected in our data.}. Assuming circular orbits, the mass of the perturbing planet $M_{\rmn{c}}$ depends on the mass of the outer, more massive transiting planet $M_{\rmn{b}}$ and its apparent eccentricity $e_{\rm{b}}$ \citep{angladaetal10}. This results in $M_{\rmn{c}}\approx0.14$ $M_{\rmn{J}}$ in the WASP-10 system. Such a planet is expected to produce strong gravitational perturbations which should be visible as the TTV of WASP-10b. To check this scenario, we generated 
synthetic $O-C$ diagrams for WASP-10b in systems with a second planet. Parameters of the transiting planet and its host star were taken from \citet{johnson09}. Initial circular and coplanar orbits were assumed for both planets. The perturbing planet was put in an orbit with the semi-major axis between 0.0218 and 0.0257 au ($\pm0.0020$ au away from the 2:1 orbital resonance). Calculations were performed using the \textsc{Mercury} package \citep{chambers99} employing the Bulirsch--Stoer integrator. Simulations covered 270 periods of WASP-10b, i.e. the time span covered by observations. No configuration was found to reproduce periodicity close to $P^{\rm{ttv}}_{1}$ and $P^{\rm{ttv}}_{2}$. The procedure was repeated for inner 3:1 and 3:2 MMRs but these cases also brought negative results.

To search for configurations with an outer perturber reproducing $P^{\rm{ttv}}_1$ and $P^{\rm{ttv}}_2$, synthetic $O-C$ diagrams were generated with the \textsc{ptmet} code which is based on perturbation theory \citep{nesvornymorbidelli08, nesvorny09}. Both planets were put in initial circular orbits. The semi-major axis $a_{\rmn{c}}$ of the perturber varied between 0.0500 and 0.3000 au in steps of $0.0001$ and $0.0002$ au for $a_{\rmn{c}}\leq0.1$ and $a_{\rmn{c}}>0.1$ au, respectively. The amplitude of the TTV signal scales nearly linearly with the perturber's mass \citep{nesvornymorbidelli08}, thus $M_{\rmn{c}}$ was fixed and set equal to 0.3 $M_{\rmn{J}}$ for the preliminary identification. 
Each simulation covered 270 periods of WASP-10b. Periodic signals reproducing $P^{\rm{ttv}}_1$ and $P^{\rm{ttv}}_2$ were identified close to 5:3 and 5:2 MMRs for perturber masses of $\sim$0.1 and $\sim$0.5 $M_{\rmn{J}}$, respectively.

\begin{figure*}
 \includegraphics[width=17cm]{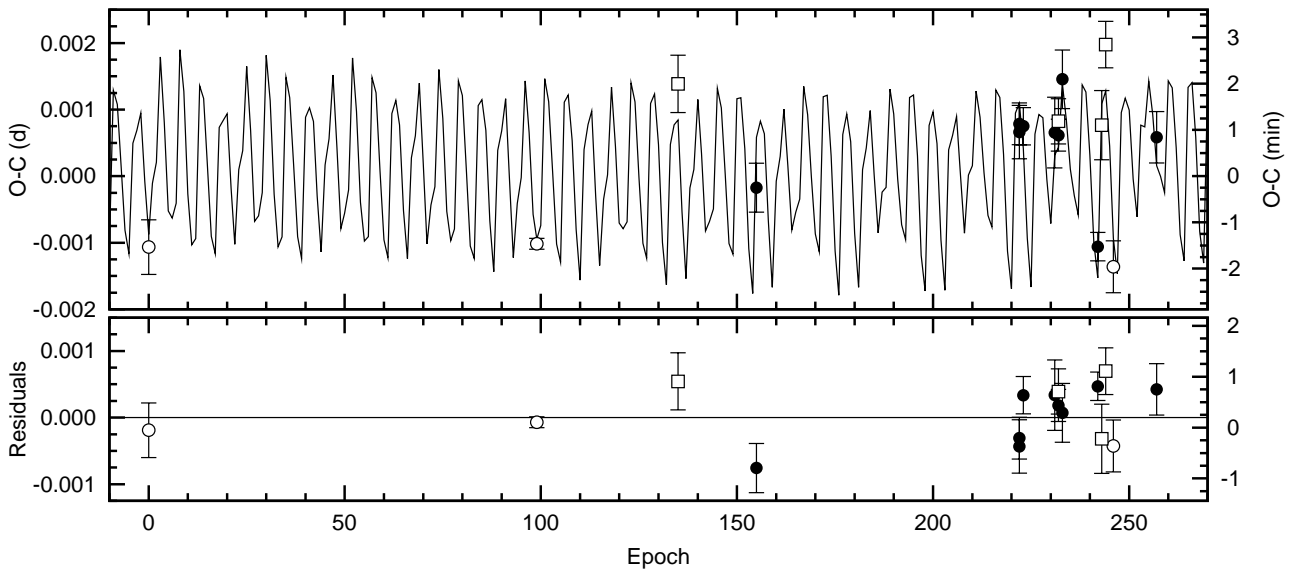}
 \caption{The $O-C$ diagram for WASP-10b with the best-fitting model (solution 1 in Table~\ref{tabela4}). The mass of the perturber is 0.10 $M_{\rmn{J}}$ and its orbital period is $\sim$5.23 d. Symbols are the same as in Fig.~\ref{rys2}. Residuals are plotted in the bottom panel. }
 \label{rys4}
\end{figure*}

To refine parameters of a perturbing body, the synthetic $O-C$ diagrams were calculated with the \textsc{Mercury} code and the Bulirsch--Stoer integrator. The initial parameters of planet c were taken from previous simulations and then refined with the mass varying in steps of 0.05 $M_{\rmn{J}}$. Each simulation covered 31000 days (i.e. $\sim$10000 periods of WASP-10b). The synthetic $O-C$ diagrams were directly fitted to observed data by shifting along the time axis and adjusting to the time span covered by observations. The best-fitting solutions are collected in Table~\ref{tabela4}. The chi-square test favours solution 1 which reproduces $P^{\rm{ttv}}_1$ with minimal $\chi^2_{\rm{red}} = 1.5$. This value is significantly smaller than $\chi^2_{\rm{red}}$ obtained in an analogous way for remaining configurations. The $O-C$ diagram with solution 1 is presented in Fig.~\ref{rys4} where the residuals are also plotted.

\begin{table}
\centering
\begin{minipage}{84mm}
\caption{Outer-perturber solutions which reproduce the observed $O-C$ variation. $P^{\rm{ttv}}$ indicates which periodicity in the $O-C$ diagram ($P_1^{\rm{ttv}}$ or $P_2^{\rm{ttv}}$) is reproduced by a solution, $a_{\rmn{c}}$ denotes the semi-major axis of the perturbing planet, $M_{\rmn{c}}$ is its mass, $P_{\rmn{c}}$ is its orbital period, $K_{\rmn{c}}$ is the expected semi-amplitude of the radial-velocity variation and $\chi^2_{\rm{red}}$ is the lowest value of reduced chi-square for direct model fitting.} 
\label{tabela4}
\begin{tabular}{l c c c c c c}
\hline
No. & $P^{\rm{ttv}}$ & $a_{\rmn{c}}$ & $M_{\rmn{c}}$  & $P_{\rmn{c}}$ & $K_{\rmn{c}}$ & $\chi^2_{\rm{red}}$ \\
    &                & (au)          & $(M_{\rm{J}})$ & (d)           & (m s$^{-1}$)  & \\
\hline 
1 & 1 & 0.0536 & 0.10 & 5.2293 & 14.2 & 1.5\\
2 & 2 & 0.0539 & 0.10 & 5.2647 & 14.1 & 2.5\\
3 & 2 & 0.0682 & 0.55 & 7.4962 & 69.1 & 2.8\\
4 & 1 & 0.0686 & 0.55 & 7.5677 & 68.9 & 2.8\\
\hline
\end{tabular}
\end{minipage}
\end{table}

\section{Spectroscopic reanalysis and stellar activity}

\begin{figure}
 \includegraphics[width=84mm]{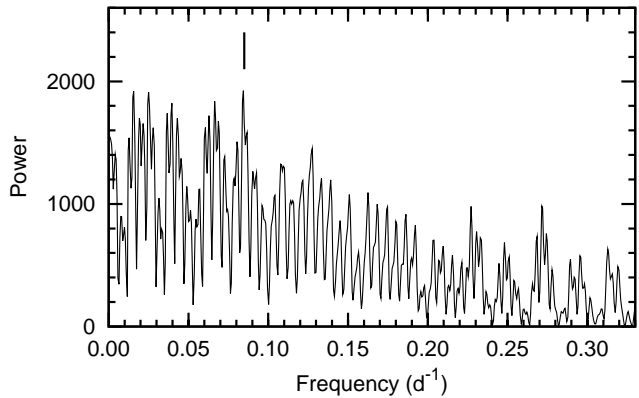}
 \caption{The Lomb--Scargle periodogram for the radial velocity residuals after removing the transiting planet in a circular orbit. A peak corresponding to the stellar rotation is marked.}
 \label{rys5}
\end{figure}

To re-assess the possible solutions, we reanalysed radial velocity (RV) data published by \citet{Christian09}. We used the \textsc{Systemic Console} software \citep{meschiarietal09}. The host star is known to be variable due to stellar rotation and starspots \citep{Smith09}. It was shown that this effect may generate the RV variations up to a few hundred m~s$^{-1}$ \citep{Hatzes02} and mimic a planetary companion on a Keplerian orbit \citep{Desort07}. We reanalysed RV data assuming WASP-10b has a circular orbit. This resulted in a best-fitting model with $\chi^2_{\rm{red}}=3.9$ and $rms=49.6$ m~s$^{-1}$.  
The periodogram of the residuals (Fig.~\ref{rys5}) reveals a periodicity of $\sim$11.84 d -- a value close to the period of stellar rotation $P_{\rm{rot}} = 11.91 \pm 0.05$ d determined by \citet{Smith09}. The fitting procedure was repeated with an additional sinusoid signal of period $P_{\rm{rot}}$ and a floating phase and amplitude. The eccentricity of WASP-10b, $e_{\rm{b}}$, was also allowed to vary. This resulted in a significantly improved model with $\chi^2_{\rm{red}}=2.5$ and $rms=31.4$ m~s$^{-1}$. The eccentricity was found to be $e_{\rm{b}}=0.013 \pm 0.063$ and it is statistically indistinguishable from zero. Therefore, we adopted $e_{\rm{b}}=0.0$ in further analysis. This result is not surprising as tidal interactions with the host star are expected to circularise the orbit of the planet \citep{zahn77}. 

\begin{figure}
 \includegraphics[width=84mm]{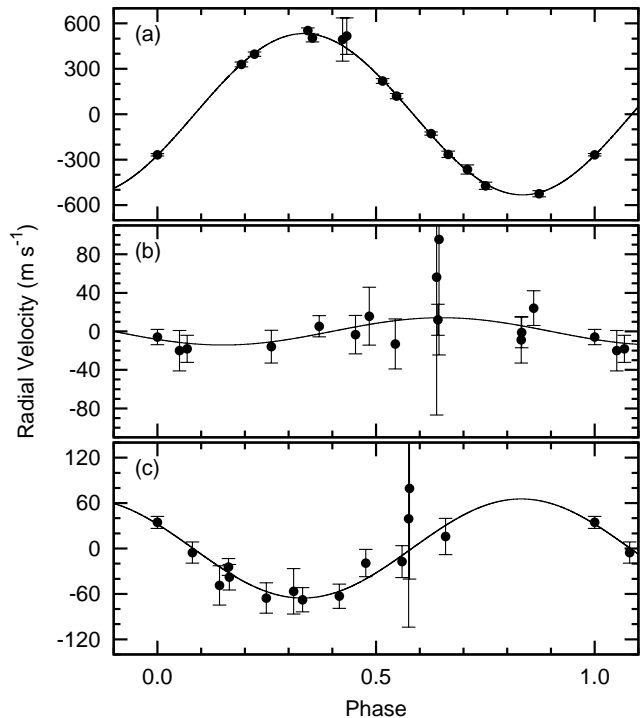}
 \caption{Radial-velocity curves after removing remaining 2 components for a) WASP-10b, b) WASP-10c and c) starspots effect. Data were taken from \citet{Christian09}. Two points with large error bars were obtained in poor weather conditions and may exhibit systematic errors due to scattered moonlight (see e.g. \citealt{Latham}).}
 \label{rys6}
\end{figure}

To check if the perturbing planet may be hidden in the RV residuals of the model composed of the planet b and starspots, the second planet with physical and orbital parameters as in solutions 1--4 was put into the system. A circular Keplerian orbit of the second planet and coplanarity were assumed for simplicity. We noted that only solutions 1 and 2 significantly decrease the $rms$ of the RV model to 24.7 and 22.5 m~s$^{-1}$, respectively. Combining results of the TTV and RV analysis, we found solution 1 the most likely. The RV curves are plotted in Fig.~\ref{rys6} for individual components, i.e. WASP-10b, WASP-10c and the starspots effect.

The RV semi-amplitude generated by stellar rotation was found to be $\sim$65 m~s$^{-1}$. RV observations by \citet{Christian09} span from August 2007 to February 2008 and are partially covered by photometric observations from the SuperWASP survey \citep{Pollacco06}. \citet{Smith09} determined the amplitude of photometric variation in the 2004 and 2006 seasons only. To find this variation in the 2007 season and verify determinations from \citet{Smith09}, the analysis of variance method (ANOVA, \citealt{schwarzenberg96}) was applied to the WASP-10 light curve divided into individual seasons. A periodic signal was identified. Then, a sinusoid was fitted to the phased light curve to determine the amplitude. We determined peak-to-peak amplitudes of $20.8\pm1.3$, $10.4\pm1.0$ and $8.5\pm0.9$ mmag for the 2004, 2006 and 2007 seasons, respectively. Our values are similar to the results of \citet{Smith09} who reported 20.2 and 12.6 mmag for 2004 and 2006, respectively. \citet{Desort07} showed that for a G2V-type star with $v \sin i = 7$ km~s$^{-1}$ and a photometric amplitude of 1 per cent (parameters close to those of WASP-10), a spot may produce RV variations with the semi-amplitude of $\sim$70 m~s$^{-1}$. This estimation is consistent with the value determined by us from the RV model analysis. 

Stellar spots may affect the transit depth (e.g. \citealt{Czesla}). We found in our data that transit depth variation is smaller than the level of 3$\sigma$ for individual transits which is thus insignificant. If stellar spots are located along a planet's path projected onto a stellar disk, the shape of the transit may be deformed. This effect may have an influence on determining mid-transit times and result in a spurious periodic TTV signal (e.g. \citealt{Alonso}). Visual inspection of photometric residuals (Fig.~\ref{rys1}) does not reveal features which could be related to stellar inhomogeneities rather than to red noise. In particular, there are no deviations visible in light-curve residuals during ingress or egress. Such deformations caused by occulting spots located close to a limb of the stellar disk are expected to generate significant TTV amplitude, even greater than a minute \citep{milleretal08b}. Furthermore, no peak close to the rotational period of the star was found in the periodogram of the $O-C$ diagram (Fig.~\ref{rys3}) and $f^{\rm{ttv}}_{1}$ is not an alias of the rotational frequency of the host star \citep{Alonso}.

The proposed model of the WASP-10 system (solution 1) was found to be dynamically stable. Simulations were performed with the \textsc{Systemic Console} employing Bulirsch--Stoer integrator and covered $10^6$ yr with a maximal time step of $10^{-4}$ yr. The eccentricity of WASP-10b was found to be close to zero, while eccentricity of WASP-10c did not exceed 0.05 with a median value of 0.024.

\section{Age of the system}

Given the relatively short rotation period of $\sim$12 days of WASP-10, gyrochronology \citep{Barnes07} can roughly yield the age of the star: For a K5 dwarf with $(B-V)=1.15$ mag \citep{schmidtkaler}, we get $270 \pm 80$ Myr only, suggesting that the star (and the planetary system) is quite young. \citet{Christian09} estimated the rotational age to be between 600 Myr and 1 Gyr by comparing the spin period of WASP-10 to stars in Hyades \citep{Terndrup}. However, one must note that according to Fig. 7 and 8 in \citet{Terndrup}, there are also Pleiades members of spectral type around K5 (like WASP-10) with $v \sin i$ around 6 km/s or lower (as WASP-10, \citealt{Christian09}). This method shows that Pleiades-like young age or an age intermediate between Pleiades and Hyades is possible for WASP-10.

To confirm the relative young age of WASP-10, we investigated its kinematic properties. The proper motion of WASP-10 is $\left(\mu_\alpha\cos\delta, \mu_\delta\right)=\left(21.4\pm2.0, -28.9\pm1.0\right)\,\mathrm{mas\,yr^{-1}}$ \citep{2003yCat.1289....0Z} and its radial velocity is $-11.44\pm0.03\,\mathrm{km\,s^{-1}}$ \citep{Christian09} yielding a heliocentric velocity of $\left(U,V,W\right)=\left(-0.3\pm0.9,-17.3\pm1.7,-8.1\pm3.2\right)\,\mathrm{km\,s^{-1}}$. First, we compared the stars's spatial velocity with these of known nearby moving groups including the Pleiades and Hyades (\citealt{2008A&A...490..135A}, e.g. their Figure 5, Table 3; \citealt{2009ApJ...692L.113Z}, e.g. their Figure 2, Table 1). We found that WASP-10 cannot be associated with any of the moving groups. In fact, it lies in an area in the $U$-$V$ plane where the star density is rather low, which may indicate that this star was either formed in isolation (very
rare) or was ejected from its parent association.

If WASP-10 was ejected recently, then it might be possible to identify its parent cluster. As stars are ejected most likely soon after cluster formation (due to a supernova event in a binary system, e.g. \citealt{1961BAN....15..265B} or via dynamical interactions in a dense cluster, e.g. \citealt{1967BOTT....4...86P}), the age of WASP-10 should be its kinematic age, i.e. it left its parent cluster some hundred million years ago. For such a long time, it is hardly possible to reconstruct the past trajectory of the system as well as the past trajectory for a potential parent cluster because both, the star and the cluster, have experienced a complete cycle (or even more) around the Galactic centre. Considering a time span up to $1\,\mathrm{Gyr}$, even periodic appearances of close encounters between the star and a cluster could occur. Moreover, it is not unlikely that WASP-10 experienced some perturbations of its path. For those reasons, the parent cluster of WASP-10 might remain unidentified if the star was ejected relatively soon after formation. If, for some reason, WASP-10 was ejected recently, finding its parent cluster could be possible. The system's peculiar spatial velocity is 
$10\pm2\,\mathrm{km\,s^{-1}}$ (the heliocentric velocity was corrected for Solar motion and Galactic rotation applying the LSR from 
\citet{2010arXiv1007.4883T} and a value of $V_{\odot,rot}=225\,\mathrm{km\,s^{-1}}$). Considering a typical cluster radius of a few parsecs, the parent cluster of WASP-10 is well identifiable if the star system was ejected not earlier than some million years ago. During 20 million years, the star travelled about $200\,\mathrm{pc}$. Taking into account the current distance to the Sun of about $100\,\mathrm{pc}$ and considering also previous age constraints, we selected open clusters within $300\,\mathrm{pc}$ from the Sun having ages between $100$ and $800\,\mathrm{Myr}$
(corresponding to the ages of the Pleiades and Hyades, respectively) from the \citet{2010yCat....102022D} database for open clusters with 
full kinematics available with the catalogue as well as the Hyades cluster (WEBDA\footnote{http://www.univie.ac.at/webda/webda.html}, \citealt{2003A&A...410..511M}), 14 clusters in total. We calculated the past orbits of WASP-10 and each cluster using the epicycle approximation \citep{1959...Lindblad,1982lbg6.conf..208W}. Then, we performed a Monte-Carlo simulation varying the distance, proper motion and radial velocity within their confidence intervals to account for their uncertainties. The goal was to find close encounters between WASP-10 and any cluster in the past. Two clusters for which close encounters with WASP-10 were found 
(within three times the cluster radius corresponding to approximately $3\sigma$), are listed in Table~\ref{tab:closeencounters}. The cluster radius, the time of the encounter and the age of the clusters are given. If WASP-10 really originated from either Platais 2 or ASCC 123, the star is about $260$ to $400\,\mathrm{Myr}$ old, in agreement with our age estimate of $270\pm80\,\mathrm{Myr}$. It is, of course, more likely that the star is ejected early in the life of a cluster. It is generally not possible to trace WASP-10 backwards reliably for the estimated age of the system.

\begin{table}
\centering
\caption{Open clusters for which close encounters with WASP-10 were found some million years in the past. The first column gives the 
cluster designation, the second and third columns indicate the separation $d$ between WASP-10 and the cluster centre and the radius $R$ of the cluster.
Column four states the time $\tau$ of the encounter (time before present) and column five quotes the cluster age as given in \citet{2010yCat....102022D}. $d$ was derived from 3D Gaussian distributions that can explain the slope of the $d$ distribution, see \citet{2010MNRAS.402.2369T}.}
\label{tab:closeencounters}
\begin{tabular}{c c c c c}
\hline
Cluster & $d$ (pc) & $R$ (pc) & $\tau$ (Myr)  &	age (Myr)\\
\hline
Platais 2 & $0\pm6$ & 10 & 13 & 400\\
ASCC 123 & $0\pm12$ & 6 & 11 & 260\\
\hline
\end{tabular}
\end{table}

The young age of the host star may explain the large radius of its planet, WASP-10b. The non-irradiated models by \citet{Baraffe03}, interpolated for such a young 3 $M_{\rmn{J}}$ planet, results in a radius of $\sim$1.2 $R_{\rmn{J}}$. Irradiation may increase the planetary radius by $\sim$10 per cent, compared to non-irradiated models \citep{Baraffe03}. An analogous procedure based on irradiated planetary models by \citet*{Fortney} and \citet*{Baraffe08} also gives a radius of $\sim$1.2 $R_{\rmn{J}}$ for a 300 Myr old giant planet. This estimate is consistent with the radius of WASP-10b reported by \citet{Christian09}, \citet{Dittmann10} and \citet{Krejcova10}. 

\section{Conclusions}

Our investigation indicates that transit timing of WASP-10b cannot be explained by a constant period of the exoplanet. The distribution of data points in the $O-C$ diagram reveals the existence of a periodic signal. We constructed a provisional hypothesis assuming the presence of the third body in the system. Combining results of 3-body simulations and RV measurements the most likely explanation of the observations is given by a model with a second planet of mass of $\sim$0.1 $M_{\rm{J}}$ and orbital period of $\sim$5.23 d. 

The reanalysis of RV data shows that WASP-10b orbits its active host star in a circular orbit. The signature of stellar rotation was found in RV measurements. This effect was misinterpreted in previous studies as a non-zero eccentricity of the transiting planet. The second planet would generate wobbling of the host star with the semi-amplitude of $\sim$14 m~s$^{-1}$ -- much less than the RV variation caused by the stellar rotation. Although we managed to distill this weak signal, we emphasis that further simultaneous spectroscopic and photometric observations are needed to study the activity of WASP-10 and verify the existence of the planet WASP-10c. 

WASP-10 was found to be a relatively young star. Assuming that this finding, which is based on the stellar gyrochronology, is correct, then the observed radius of WASP-10b is consistent with theoretical models and no tidal heating is required to match its radius. If planetary gas giants form by gravitational contraction, then young planets should be larger than older ones. If the planets of WASP-10 formed at larger separation than their present location, then migration must have taken place within the young age of the star.

The existence of WASP-10c could be independently confirmed if it is a transiter. Assuming that its radius is $\sim$0.4 $R_{\rm{J}}$, i.e. similar to exoplanets of the similar mass like HAT-P-11b \citep{hat11} or Kepler-4b \citep{kepler4}, the expected flux drop during a transit would be $\sim$0.3 per cent or even greater if we consider the young age of the system. This could be observable with a large ground-based telescope. 

In the proposed model the ratio of orbital periods of the planets $P_{\rmn{c}}/P_{\rmn{b}}=1.69$ is very close to 5:3 orbital resonance. A significant fraction of known multi-planet systems are in MMRs, mainly in low-order ones. Two planets in wide orbits and trapped in a 3:2 MMR were found around HD~45364 by RV measurements \citep{Correia09}. Planets in the 4-planetary system around GJ~581 \citep{Mayor09} are very close to 5:2 (GJ~581b and c) and 5:3 (GJ~581e and b) MMRs \citep{Papaloizou10}. Therefore, we conclude that the architecture of the WASP-10 system would not be completely unusual.

\section*{Acknowledgments}

We are grateful to the anonymous referee for remarks and inspiring us to study kinematic membership. We thank J.Dittmann for providing the original photometric data. We also thank G.Nowak and T.Eisenbeiss for discussions about stellar spots and gyrochronology. GM acknowledges support from the EU in the FP6 MC ToK project MTKD-CT-2006-042514. DD acknowledges support from the project DO 02-362 of the Bulgarian Scientific Foundation. RN would like to acknowledge general support from DFG. GM, RN and AN acknowledge support from the DAAD PPP--MNiSW project 50724260--2010/20011 \textit{Eclipsing binaries in young clusters and planet transit time variations}. AN was also supported by the Polish Ministry of Science and Higher Education grant NN 203 510938. NT acknowledges financial support from Carl-Zeiss-Stiftung. SR and CM acknowledge support from DFG in programs NE 515/32-1 and SCHR 665/7-1, respectively. TK, JB and VK acknowledge support from the Marie Curie International Reintegration Grant FP7-200297, grant GA \v{C}R GD205/08/H005, the National scholarship programme of Slovak Republic and partly from VEGA 2/0078/10, VEGA 2/0074/09. A part of this paper is a result of PAN/BAN exchange and joint research project \textit{Spectral and photometric studies of variable stars}. We have used Exoplanet Transit Database, http://var.astro.cz/ETD and data from the WASP public archive in this research. The WASP consortium comprises of the University of Cambridge, Keele University, University of Leicester, The Open University, The Queen's University Belfast, St. Andrews University and the Isaac Newton Group. Funding for WASP comes from the consortium universities and from the UK's Science and Technology Facilities Council.

\bsp

\label{lastpage}

\end{document}